\documentclass[acmsmall,screen]{acmart}

\setcopyright{none}
\renewcommand\footnotetextcopyrightpermission[1]{}
\acmConference[FSE 2027 Draft]{ACM International Conference on the Foundations
of Software Engineering}{July 12--16, 2027}{Shenzhen, China}

\usepackage{amsmath,amsfonts}
\usepackage{graphicx}
\graphicspath{{./}}
\usepackage{booktabs}
\usepackage{array}
\usepackage{multirow}
\usepackage{xcolor}
\usepackage{enumitem}

\setlist[itemize]{leftmargin=*}
\setlist[enumerate]{leftmargin=*}

\usepackage{listings}
\lstdefinelanguage{Verus}{
  morekeywords={fn,spec,proof,requires,ensures,decreases,invariant,assert,
    forall,exists,int,bool,i8,let,mut,if,else,return,use,pub,old,by,
    recommends,returns,true,false},
  sensitive=true,
  morecomment=[l]{//},
  morestring=[b]",
}
\newlist{researchquestions}{enumerate}{1}
\setlist[researchquestions,1]{
  label=\textbf{RQ\arabic*},
  leftmargin=*,
  topsep=2pt,
  itemsep=2pt
}

\newcommand{\finding}[2]{%
  \par\addvspace{\smallskipamount}\noindent
  \fbox{\parbox{\dimexpr\columnwidth-2\fboxsep-2\fboxrule\relax}{%
    \small\textbf{Finding #1.} #2}}%
  \par\addvspace{\smallskipamount}%
}

\AtBeginDocument{%
}

\newcommand{\xzw}[1]{\textcolor{red}{[XZW: #1]}} %added by xzw

\begin{document}

% Working title. Revisit after the four RQ findings are frozen.
\title{VerusEval: Evaluating LLM-Generated Verus Specifications Beyond Verifier Acceptance}

% The anonymous option hides author information in the review PDF. Replace this
% block only when preparing a non-anonymous or camera-ready version.
\author{Ding Wang}
\orcid{0009-0008-2582-5224}
\affiliation{%
  \institution{Shenzhen University}
  \department{College of Computer Science and Software Engineering}
  \city{Shenzhen}
  \country{China}
}
\email{lifetruthuuu@gmail.com}

\author{Chao Jiang}
\orcid{0009-0008-3173-5196}
\affiliation{%
  \institution{Shenzhen University}
  \department{College of Computer Science and Software Engineering}
  \city{Shenzhen}
  \country{China}
}
\email{chaoccc22@gmail.com}

\author{Dugang Liu}
\orcid{0000-0003-3612-709X}
\correspondingauthor
\affiliation{%
  \institution{Shenzhen University}
  \department{College of Computer Science and Software Engineering}
  \city{Shenzhen}
  \country{China}
}
\email{dugang.ldg@gmail.com}

\author{Zhiwu Xu}
\orcid{0000-0001-6727-440X}
\affiliation{%
  \institution{Shenzhen University}
  \department{College of Computer Science and Software Engineering}
  \city{Shenzhen}
  \country{China}
}
\email{xuzhiwu@szu.edu.cn}

\author{Cheng Wen}
\orcid{0000-0003-1826-6213}
\affiliation{%
  \institution{Shenzhen University}
  \department{College of Computer Science and Software Engineering}
  \city{Shenzhen}
  \country{China}
}
\email{wencheng@xidian.edu.cn}

\author{Zhong Ming}
\orcid{0000-0001-9310-3460}
\affiliation{%
  \institution{Shenzhen University}
  \department{College of Computer Science and Software Engineering}
  \city{Shenzhen}
  \country{China}
}
\email{mingz@szu.edu.cn}

\author{Yuwei Liu}
\orcid{0000-0001-5170-3388}
\affiliation{%
  \institution{Ant Group}
  \department{Ant Research}
  \city{Hangzhou}
  \country{China}
}
\email{lyw458372@antgroup.com}

\author{Xinyi Wan}
\orcid{0009-0004-3543-499X}
\affiliation{%
  \institution{Ant Group}
  \department{Ant Research}
  \city{Hangzhou}
  \country{China}
}
\email{wanxinyi.wxy@antgroup.com}

\author{Lin Huang}
\orcid{0009-0002-5659-1471}
\affiliation{%
  \institution{Ant Group}
  \department{Ant Research}
  \city{Hangzhou}
  \country{China}
}
\email{linyu.hl@antgroup.com}

\renewcommand{\shortauthors}{D. Wang, C. Jiang, D. Liu, Z. Xu, C. Wen, Z. Ming, Y. Liu, X. Wan, and L. Huang}

\begin{abstract}
Large language models (LLMs) are increasingly used to generate Verus contracts for Rust programs together with the proofs that establish them.
Yet verifier acceptance, the common measure of success, shows only that the implementation satisfies the generated contract, \iffalse and how far it overstates contract quality remains poorly understood.\fi so it leaves open whether the contract captures the intended behavior. Assessing specification generation therefore requires an evaluation methodology, not a single metric. 
We present VerusEval, a framework that evaluates generated contracts along six dimensions, including Verus-native checks for semantic triviality, validated input--output (I/O) suites, and contract equivalence, and that relates them so that the strength and blind spots of each become measurable.
Evaluating 13,716 artifacts from four workflows, five LLMs, and two prompting settings on 762 tasks, we find that acceptance overstates quality: only 42.3\% of the accepted artifacts complete all three
semantic checks without a flag,
and across workflows and models, higher acceptance does not consistently come with more correct contracts.
I/O evaluation rejects 23.6\% of the accepted artifacts, yet only 63.2\% of those that pass it are proved equivalent. 
Of the accepted artifacts flagged by contract comparison, 84.1\% have a rejected completeness obligation,
and the direction of a failure prioritizes inspection of the generated contract or the reference; it does not by itself establish which artifact is defective.
Screening and review further identify 36 candidate defects among 61 selected reference contracts, and repairing one of them makes 10 of its 11 failing artifacts equivalent.
Based on these findings, we derive four actionable lessons for evaluation, generation, and benchmark maintenance.
\end{abstract}


\maketitle

\section{Introduction}
\label{sec:introduction}
Formal verification establishes that a program satisfies a mathematical specification, providing guarantees about the behaviors covered by that specification.
Its use in verified kernels and distributed systems demonstrates its value for building reliable software~\cite{klein2009sel4,gu2016certikos,hawblitzel2015ironfleet}.
For Rust, Verus~\cite{lattuada2023verus} supports deductive verification of executable code annotated with specifications and proof code, and it discharges the resulting proof obligations with an SMT solver.
Developers express admissible inputs and output guarantees through function contracts~\cite{meyer1992contract}, but writing suitable contracts and the proofs needed to establish them requires substantial expertise and effort~\cite{hahnle2019deductive}.
Large language models (LLMs) can automate specification and proof development for Rust.
AlphaVerus~\cite{aggarwal2025alphaverus} generates code and contracts through translation and verifier feedback, while StarVerus~\cite{jiang2026starverus} coordinates multiple agents to generate Verus specifications and proofs.
Related workflows such as AutoVerus~\cite{yang2025autoverus} and VeruSAGE~\cite{yang2025verusage} generate and repair proofs for supplied contracts.
These advances make it increasingly feasible to produce verified Rust artifacts automatically. 
%and create a need to assess the quality of the specifications they contain.

\iffalse When an LLM generates the contract as well as the proof, the contract becomes part of the output, and verification checks the implementation against it but not the contract against the intended behavior.
Specification generation studies nevertheless measure success mainly by whether the generated contracts verify~\cite{ma2025specgen,jiang2026starverus}.
Acceptance establishes only that the implementation satisfies the generated contract, so a contract that omits a required input constraint or output guarantee, excludes valid inputs, or is vacuous can still verify.
Counting such artifacts as successes overstates what generation achieves and can mislead comparisons between workflows and models.\fi 

Once the contract is generated rather than supplied, the bottleneck moves from generation to evaluation: verification checks the implementation against the contract, and nothing checks the contract against the intended behavior. 
A verified artifact is therefore only as strong as its contract, so a contract that omits a required constraint, excludes valid inputs, or is vacuous still passes verification. 
Acceptance consequently becomes an unreliable basis for comparison: \iffalse because a weaker contract is usually easier to prove,\fi a workflow that emits looser contracts can appear more successful without delivering stronger guarantees. 
Evaluating generated contracts thus requires evidence beyond acceptance, but the available kinds of evidence differ in strength.
Concrete input--output (I/O) cases, which prior work uses for postconditions, Rocq specifications, and Verus specifications~\cite{endres2024postconditions,yang2026coins,agarwal2026verusspecgym}, can expose a wrong contract but cover only sampled behaviors.
Logical comparison with a reference covers all inputs, but it is only as reliable as the reference, and a rejected proof attempt does not yield a counterexample.
Text similarity and LLM judgements are cheap but cannot establish logical relations.
%It remains unclear how far acceptance overstates contract quality, which differences each kind of evidence can reveal, how generated contracts deviate from their references, and how reliable these references are.

Recent evaluation frameworks each foreground one kind of evidence on their own artifacts and tasks.
Verus-SpecGym~\cite{agarwal2026verusspecgym}, Coins~\cite{yang2026coins}, CodeSpecBench~\cite{chen2026codespecbench}, and TLA$^+$-Bench~\cite{bisharat2026tlabench} ground their evaluation in execution. Clover~\cite{sun2024clover} checks consistency among code, docstrings, and annotations. A study~\cite{prasetya2026talkischeap} of postcondition formalization reports that generated tests expose solutions otherwise counted correct, and LiveFMBench~\cite{xu2026livefmbench} shows that prover gaming inflates measured accuracy.
These studies establish that acceptance overstates contract quality but leave open three questions for an evaluation methodology.
How strong is each kind of evidence, and where is it blind, when all are applied to the same artifacts?
Can a difference from a reference become a diagnosis that indicates which contract to revise?
Who audits the reference on which the other evidence depends?

We introduce \emph{VerusEval}, a framework for systematically evaluating LLM-generated Verus specifications.
It organizes these kinds of evidence into six dimensions with different roles.
Text similarity and intent judgement give lightweight signals of how a contract is written and what it appears to intend.
Verifier validity records how far an artifact progresses through verification, and semantic triviality detects vacuous contracts without a reference.
Specification correctness uses proof obligations discharged by Verus itself to check whether a contract is sound (equivalent to or stronger than the reference) and complete (equivalent to or weaker).
Behavior correctness checks the contract on a validated I/O suite. \iffalse whose cases are labeled by the reference\xzw{shall we remove the relation to reference?}.\fi 
When a comparison with the reference fails, the rejected obligation shows the direction of the difference, which prioritizes inspection of the generated contract or the reference but does not by itself establish which artifact is defective.
We apply VerusEval to 13,716 generated artifacts over 762 tasks, covering 18 configurations of four workflows, five models, and two prompting settings.
The study includes StarVerus and adaptations of AlphaVerus, AutoVerus, and VeruSAGE for contract generation.
Our contributions are:
\begin{itemize}
  \item \textbf{A specification evaluation framework.}
  We define and implement six dimensions for evaluating generated Verus contracts. 
  Four are diagnostic, and three of them adapt established checks: vacuity detection~\cite{beer2001vacuity}, the classical contract refinement order~\cite{meyer1992contract,liskov1994behavioral}, and finite-sample I/O testing~\cite{endres2024postconditions,molina2021evospex}. \iffalse rather than introducing a new checking principle.\fi  
  Our contribution is to instantiate them natively in Verus and to calibrate them on the same artifacts, so that the strength and blind spots of each become measurable. 
  Encoding soundness and completeness as native proof obligations makes the direction of a deviation readable from the failed obligation.
  %including native Verus checks for semantic triviality and contract comparison and a validated I/O suite with correct pairs, wrong-output pairs, and invalid inputs.
  
  \item \textbf{A systematic empirical study.}
  We quantify how well verifier acceptance reflects contract quality and compare generation workflows, models, and prompting settings.
   Only 42.3\% of accepted artifacts complete the triviality, I/O, and contract-comparison checks without a flag, and although acceptance is positively associated with quality at the artifact level, across workflows and models, higher acceptance does not consistently come with higher specification correctness. 
  \iffalse Acceptance carries information about quality but overstates it as a success criterion (Finding 1), and it does not reliably rank configurations (Finding 2). \fi
  
  \item \textbf{An analysis of evaluation evidence, failure modes, and reference quality.}
  I/O evaluation rejects 23.6\% of the accepted artifacts, yet only 63.2\% of those that pass it are proved equivalent to their references. \iffalse and among 943 controlled contract variants, the 72 that strengthen the postcondition escape the suites by construction rather than for lack of a case.\fi 
  Of the accepted artifacts flagged by contract comparison, 84.1\% have a rejected completeness obligation, that is, a failed proof attempt rather than a confirmed difference.
  Screening selects 61 of the 762 reference contracts, and review labels 36 of them as candidate defects.
  We demonstrate the resulting audit–repair procedure on one reference: restoring its missing guarantee makes 10 of its 11 failing artifacts equivalent.
  These results inform practical lessons for generation, evaluation, and benchmark maintenance.
\end{itemize}

\section{Background}
\label{sec:background}
\subsection{Verus}
\label{sec:verus}
\label{sec:establishes}

Verus~\cite{lattuada2023verus} is a deductive verifier for Rust: developers annotate functions with contracts and proof code, and an SMT solver discharges the resulting proof obligations automatically.
Functions come in three modes: \texttt{exec} functions are compiled Rust, \texttt{spec} functions give pure mathematical definitions for contracts to use, and \texttt{proof} functions package reusable lemmas.
An \texttt{exec} function's contract consists of a \texttt{requires} clause, which defines its admissible inputs, and an \texttt{ensures} clause, which states what the function guarantees on return.
Contracts, together with the \texttt{spec} functions they call, form the specification of a program.
The proof code instead serves as scaffolding: assertions, loop invariants, decreases clauses, ghost variables, and proof lemmas help the verifier discharge the proof obligations that the contract and implementation induce.
A file reaches acceptance by clearing three successive stages: parsing, frontend checking (including name resolution, typing, and Verus mode rules), and proof.
Verus accepts a file only when every function in it verifies.
Acceptance certifies that the executable implementation satisfies the contract supplied in the same file, given the supplied proof code.

\subsection{Satisfiability, Soundness, and Completeness}
\label{sec:semantics}

Let $C$ denote a Verus function's executable implementation, and let $P$ and $Q$ denote the precondition and postcondition expressed by its \texttt{requires} and \texttt{ensures} clauses, respectively.

The Hoare triple $\{P\}C\{Q\}$~\cite{hoare1969axiomatic} holds if, for any input $I$ that satisfies the precondition $P$ and on which $C$ terminates, the resulting output $O$ satisfies the postcondition $Q$. However, verifier acceptance, which establishes that $\{P\}C\{Q\}$ holds, is insufficient as a sole evaluation metric. For instance, if $P$ is unsatisfiable (i.e., always false) or $Q$ is a tautology (i.e., always true), then \emph{any} program $C$ trivially satisfies $\{P\}C\{Q\}$, rendering the verification result vacuous.

We say that a program $C$ satisfies a specification $(P,Q)$ if $\{P\}C\{Q\}$ holds. Likewise, an input--output pair $(I, O)$ is accepted by $(P,Q)$ if $O$ satisfies $Q$ whenever $I$ satisfies $P$. In terms of program behavior, $C$ satisfies $(P,Q)$ if every correct input--output pair $(I, O)$ of $C$ is accepted by $(P,Q)$.

This captures only one direction, referred to as \textit{completeness}: it ensures that no correct behaviors of $C$ are excluded from the specification. The complementary direction, referred to as \textit{soundness}, requires that every input--output pair $(I, O)$ accepted by the specification $(P,Q)$ must itself be a correct input--output pair of $C$. Intuitively, soundness guarantees that all behaviors characterized by the specification are indeed exhibited by the program.

An ideal specification $(P, Q)$ should be both sound and complete. If it is, then, by contraposition, every pair that $C$ cannot produce is rejected by $(P, Q)$, and we call such pairs incorrect.  Following prior work on evaluating postconditions~\cite{endres2024postconditions,molina2021evospex}, we evaluate $(P, Q)$ against sets of observed correct and sampled incorrect input--output pairs, checking whether it accepts the former and rejects the latter. These checks provide finite-sample evidence for soundness and completeness.

Such finite sets, however, may not fully represent the program's complete behavioral space. In certain benchmarks, a reference specification $(P_0, Q_0)$ is provided, which we assume faithfully captures the intended program semantics\footnote{A provided specification in practice is always more abstract than any implementation that satisfies it, and therefore admits a broader set of behaviors.}. Relative to this reference, a generated specification $(P, Q)$ is \emph{sound} (i.e., equivalent to or stronger than) if every input--output pair accepted by $(P,Q)$ is accepted by $(P_0,Q_0)$. It is \emph{complete} (i.e., equivalent to or weaker than) if every input--output pair accepted by $(P_0,Q_0)$ is accepted by $(P,Q)$. It is \emph{equivalent} to $(P_0, Q_0)$ iff it is both sound and complete.
\iffalse Relative to this reference, a generated specification $(P, Q)$ is \emph{sound} (i.e., not weaker than the reference) if $P_0 \Rightarrow P$ and $Q \Rightarrow Q_0$; it is \emph{complete} (i.e., not stronger than the reference) if $P \Rightarrow P_0$ and $Q_0 \Rightarrow Q$. A generated specification $(P, Q)$ is \emph{equivalent} to the reference specification $(P_0, Q_0)$ if and only if it is both sound and complete with respect to $(P_0, Q_0)$.
The four implications instantiate the classical refinement order on contracts~\cite{meyer1992contract,liskov1994behavioral}: with the reference now in the role of the program, soundness means every pair accepted by $(P,Q)$ is accepted by $(P_0,Q_0)$, for which $P_0\Rightarrow P$ and $Q\Rightarrow Q_0$ suffice, and completeness means the converse. 
\fi 

\begin{figure}[htbp]
\centering
\includegraphics[width=\linewidth]{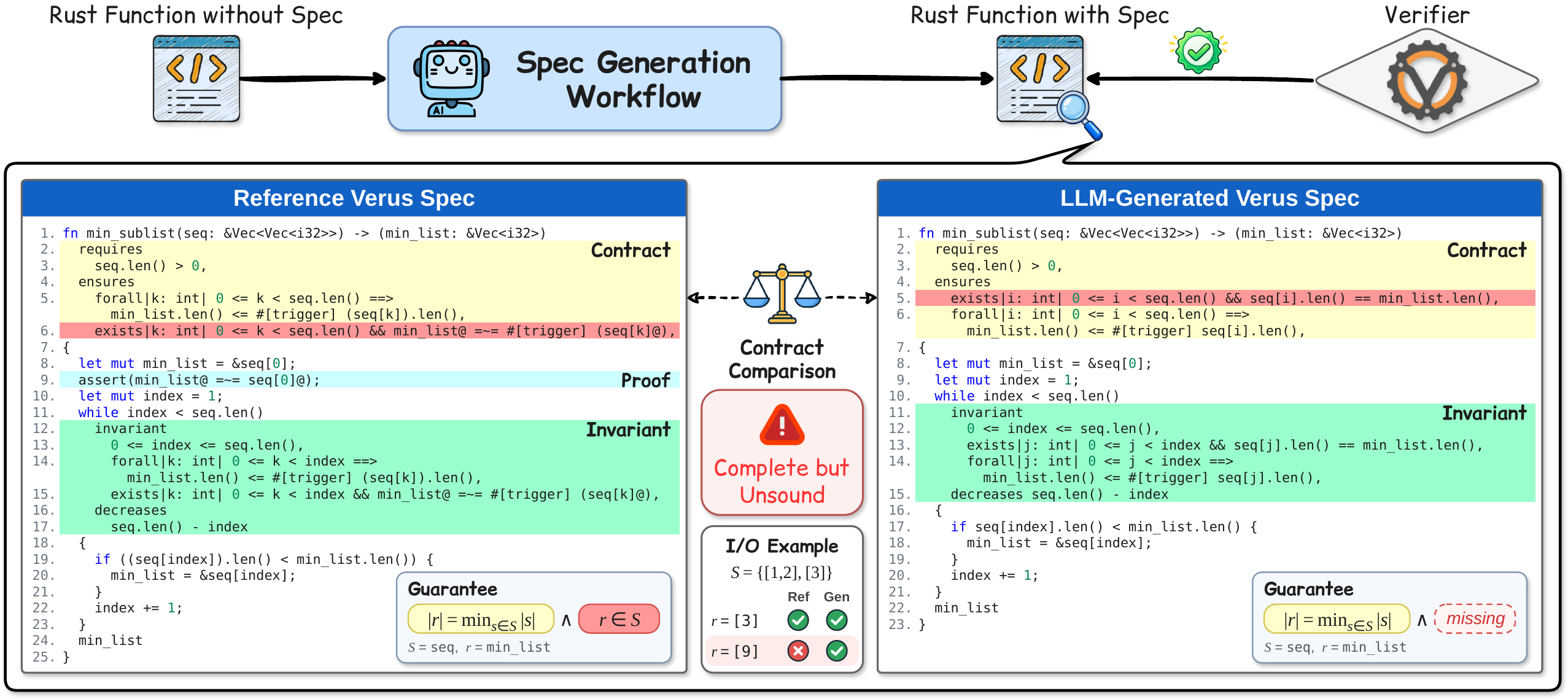}
\caption{A contract can verify while omitting an output guarantee.}
\Description{A Rust function without a specification passes through an LLM-based specification generation workflow, and the verifier accepts the resulting Rust function with its specification. A zoomed view shows the reference and the LLM-generated Verus code for min\_sublist, with contract, invariant, and proof lines highlighted. Both contracts require a nonempty collection seq and ensure that the returned list min\_list is no longer than any member. The reference additionally ensures that min\_list equals some member of seq, whereas the generated contract only ensures that some member has the same length. A guarantee box in the lower right of each listing summarizes the contract, writing S for seq and r for min\_list: the reference guarantees that r has the minimum length and is a member of S, whereas the generated box keeps the minimum length and marks the membership guarantee as missing. Contract comparison concludes that the generated contract is complete but unsound. An I/O example with S containing the lists [1,2] and [3] shows that both contracts accept r = [3], whereas only the generated contract accepts r = [9].}
\label{fig:motivating-case}
\end{figure}

Figure~\ref{fig:motivating-case} shows \texttt{min\_sublist}, which returns a shortest list from a nonempty collection $S$.
Its generated contract constrains the returned list $r$ to have the minimum length but omits $r\in S$.
For $S=\{[1,2],[3]\}$, this contract permits $r=[9]$, so it is unsound relative to the reference even though the artifact verifies.

\section{The VerusEval Framework}
\label{sec:framework}
\begin{figure}[htbp]
\centering
\includegraphics[width=\linewidth]{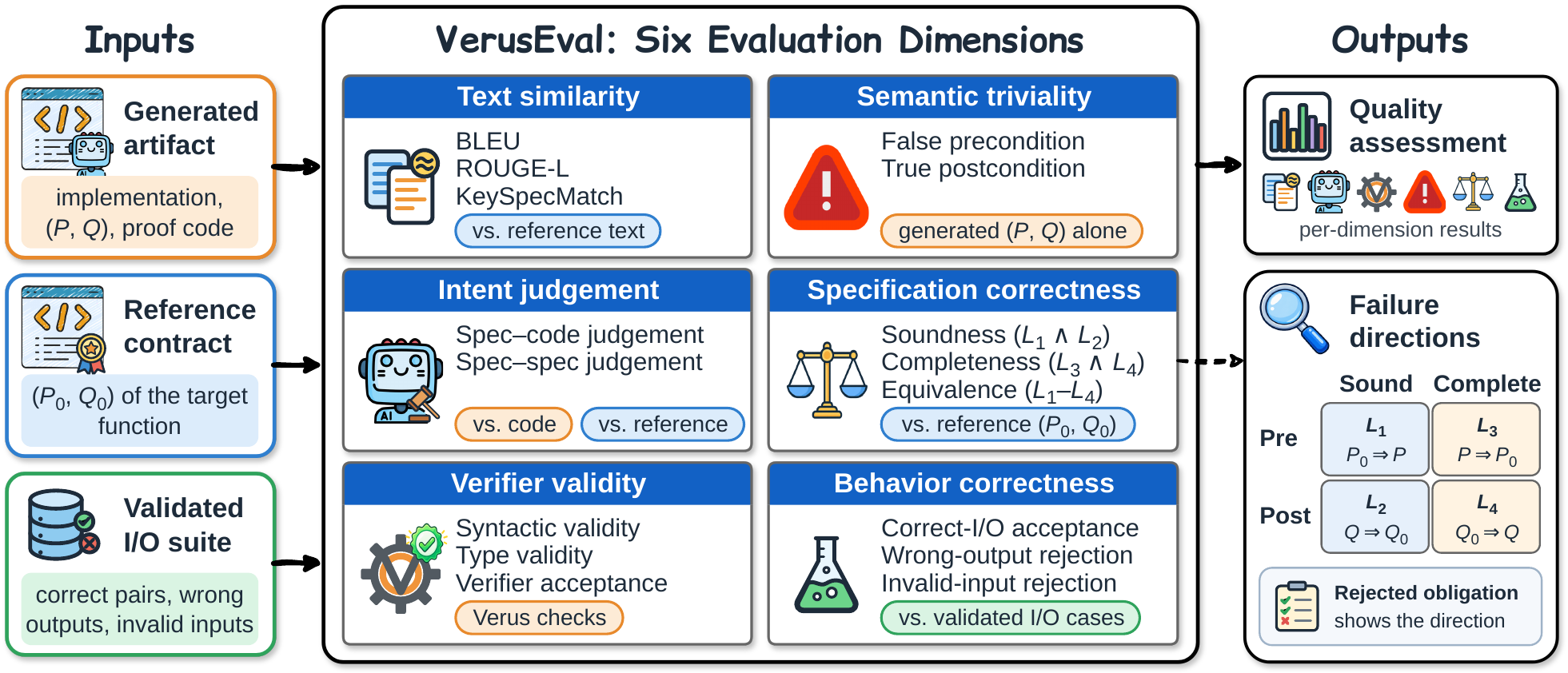}
\caption{Overview of VerusEval.}
\Description{A generated artifact, a reference contract, and a validated I/O suite feed six evaluation dimensions: text similarity, intent judgement, verifier validity, semantic triviality, specification correctness, and behavior correctness. The outputs are per-dimension results and failure directions: L1 and L2 check soundness, L3 and L4 check completeness, and all four together check equivalence to the reference contract. A rejected obligation shows the direction of a difference.}
\label{fig:framework}
\end{figure}

Because verifier acceptance alone cannot show whether a generated contract is vacuous or fails to capture the intended behavior (Section~\ref{sec:semantics}), VerusEval also evaluates what the contract requires and guarantees.
Figure~\ref{fig:framework} outlines the evaluation.
The unit of evaluation is a generated artifact: a Verus file containing the implementation of the task's target function, a generated specification, and proof code.
VerusEval evaluates each artifact along six dimensions, some of which also draw on the task's reference contract or on a validated I/O suite.

\subsection{Evaluation Dimensions}
\label{sec:catalogue}

\begin{itemize}
  \item \textbf{Text similarity} measures how closely the generated contract resembles the reference contract in wording.
  It reflects how a contract is written rather than what it means, so logically equivalent contracts can still differ.
  \item \textbf{Intent judgement} uses an LLM to assess whether the generated contract is consistent with the implementation or the reference contract.
  Its explanations help locate potential problems.
  \item \textbf{Verifier validity} records how far the artifact progresses through verification (Section~\ref{sec:verus}).
  Acceptance leaves open whether the contract captures the required behavior.
  \item \textbf{Semantic triviality} detects vacuous contracts~\cite{beer2001vacuity}.
  A false precondition excludes every input, and a true postcondition constrains no output.
  \item \textbf{Specification correctness} assesses the soundness and completeness of the generated contract relative to the reference.
  It relates the generated precondition and postcondition to their reference counterparts by logical implication, so its conclusions are only as reliable as the reference.
  \item \textbf{Behavior correctness} checks whether the generated artifact admits the correct cases and rejects the incorrect cases of the validated I/O suite.
  Its cases are labeled using the reference implementation and contract, so it shares the reference dependence of specification correctness.
\end{itemize}
The six dimensions play two roles. Text similarity and intent judgement are descriptive and never flag an artifact, since neither can establish a logical relation. The other four are diagnostic, and three of them flag artifacts through the triviality check, the I/O suite, and contract comparison. %Section 5.1 reports how much each contributes.

\subsection{Metric Definitions}
\label{sec:implementation}

Table~\ref{tab:metric-catalog} lists the metrics in each dimension.
We describe them in the same order.

\begin{table}[htbp]
\caption{Core metrics.}
\label{tab:metric-catalog}
\centering
\footnotesize
\setlength{\tabcolsep}{3pt}
\begin{tabular}{@{}ll@{\hspace{1.25em}}ll@{}}
\toprule
Dimension / metric & \begin{tabular}[b]{@{}l@{}}Comparison object /\\evidence\end{tabular}
  & Dimension / metric & \begin{tabular}[b]{@{}l@{}}Comparison object /\\evidence\end{tabular} \\
\cmidrule(r){1-2}\cmidrule(l){3-4}
\textit{Text similarity} & Reference contract text & \textit{Semantic triviality} & \\
\quad BLEU & & \quad False precondition & Generated precondition \\
\quad ROUGE-L & & \quad True postcondition & Generated postcondition \\
\quad KeySpecMatch & & \textit{Specification correctness} & Reference contract \\
\textit{Intent judgement} & & \quad Reference-relative soundness & \\
\quad LLM spec--code judgement & Implementation & \quad Reference-relative completeness & \\
\quad LLM spec--spec judgement & Reference contract & \quad Reference equivalence & \\
\textit{Verifier validity} & & \textit{Behavior correctness} & \\
\quad Syntactic validity & Parser diagnostics & \quad Correct-I/O acceptance & Correct input--output pairs \\
\quad Type validity & Frontend checks & \quad Wrong-output rejection & Wrong-output pairs \\
\quad Verifier acceptance & Proof outcomes & \quad Invalid-input rejection & Invalid inputs \\
\bottomrule
\end{tabular}
\end{table}

\textit{Text similarity.}
From each file, we extract the contract clauses and all \texttt{spec} function definitions, since contracts call these functions.
The three metrics compare the normalized tokens of this text between the generated and reference files.
BLEU~\cite{papineni2002bleu,post2018clarity} and ROUGE-L~\cite{lin2004rouge} measure general textual overlap and treat all tokens alike.
KeySpecMatch instead focuses on specification-specific features: clause kinds, selected Verus tokens such as quantifiers and logical operators, and identifiers.
It reports the fraction of reference features that also occur in the generated contract.

\textit{Intent judgement.}
The two judgements follow the LLM-as-a-judge setting~\cite{zheng2023judge} and differ in what the judge sees.
The spec--code judgement receives the full generated file, so the judge infers the intended behavior from the implementation.
The spec--spec judgement receives the generated and reference contracts, together with their \texttt{spec} functions, and treats the reference as the intended behavior.
Each judgement returns a score between 0 and 1, where 1 indicates full consistency and 0 indicates an unrelated, vacuous, or contradictory contract.
The judge also explains its score by listing missing constraints, overly strong constraints, and vacuous clauses.

\textit{Verifier validity.}
The three metrics are binary and follow the verification stages in Section~\ref{sec:verus}.
Syntactic validity holds if the artifact parses.
Type validity holds if the artifact also passes frontend checking.
Verifier acceptance holds if Verus also discharges every proof obligation in the file.

\textit{Semantic triviality.}
Both triviality metrics are defined by proof obligations.
We write each obligation as a Verus \texttt{proof} function and discharge it with Verus itself.
This native encoding preserves \texttt{spec} functions, quantifiers, and the distinction between initial and final states.
Each proof attempt has one of three outcomes: the obligation is \emph{proved} when Verus discharges it, \emph{rejected} when a completed attempt fails, and \emph{unresolved} when the attempt does not complete, for example because of a timeout.
Only a successful proof establishes the obligation. A rejected or unresolved attempt does not establish its negation.
Each metric checks one obligation on the generated contract alone:
\begin{align}
\text{False Precondition } (T_1)  &: \quad P \equiv \mathit{false}, \label{eq:t1}\\
\text{True Postcondition } (T_2)  &: \quad Q \equiv \mathit{true}. \label{eq:t2}
\end{align}
Here, $\equiv$ denotes logical equivalence over all values of the relevant inputs and outputs.
$T_1$ holds if the precondition is unsatisfiable, and $T_2$ holds if the postcondition is a tautology.
$T_2$ does not assume $P$, so a false precondition cannot make it hold.
A proved obligation certifies the corresponding defect.

\textit{Specification correctness.}
We check the soundness and completeness of Section~\ref{sec:semantics} directly by comparing the generated contract $(P,Q)$ with the reference $(P_0,Q_0)$.
Like $T_1$ and $T_2$, each of the following four implications is encoded as a native Verus proof obligation:
\begin{align}
\text{Stronger Precondition } (L_1) &: \quad P_0 \Rightarrow P, \label{eq:l1}\\
\text{Weaker Postcondition } (L_2) &: \quad Q \Rightarrow Q_0, \label{eq:l2}\\
\text{Weaker Precondition } (L_3) &: \quad P \Rightarrow P_0, \label{eq:l3}\\
\text{Stronger Postcondition } (L_4) &: \quad Q_0 \Rightarrow Q. \label{eq:l4}
\end{align}
The four implications instantiate the classical refinement order on contracts~\cite{meyer1992contract,liskov1994behavioral}.
When $L_1$ and $L_2$ hold, the generated contract is sound. When $L_3$ and $L_4$ hold, it is complete. When all four hold, it is equivalent to the reference. 
%Soundness requires $L_1$ and $L_2$, completeness requires $L_3$ and $L_4$, and equivalence requires all four.
We name each obligation after the difference that its rejection suggests: a rejected $L_1$, for example, points to a generated precondition that excludes inputs admitted by the reference. A rejected obligation is a failed proof attempt, not a confirmed semantic difference. Because $L_2$ and $L_4$ compare postconditions without precondition guards, their rejection can also arise outside the inputs admitted by one or both contracts. Thus, $L_1$--$L_4$ are sufficient but not necessary for the inclusions of Section~\ref{sec:semantics}, as $L_2$ and $L_4$ compare postconditions on all inputs rather than only on inputs satisfying the preconditions, making them stronger than needed. 
\iffalse and comparing postconditions on all inputs makes $L_2$ and $L_4$ stronger than needed: postconditions differing only where both preconditions exclude inputs are behaviorally equivalent and still yield a rejected implication.
A rejected obligation is therefore a failed proof attempt, not a confirmed difference.
%Section 5.5 exhibits one such case and Section 7 discusses the consequence. We keep the unconditional form because it isolates a clause.
\fi 
We keep the unconditional forms because they attribute a difference to a single clause and keep the comparison independent of the implementation.
A domain-restricted alternative such as $\forall I,O.\,I \models P_0 \land P\Rightarrow(O\models Q \Rightarrow O \models Q_0)$ is more precise but entangles both clauses in one obligation, making the failing direction harder to read.

%We still check the unconditional forms because preconditions constrain the initial state and postconditions the final state, yet preconditions need not hold after execution.  
%Restricting comparison to the common domain then requiring quantifying over output states, yielding $\{P\land P_0\}\,C\,\{Q\Rightarrow Q_0\}$, which makes the contract-comparison metric depend on the implementation, and thus varying across artifacts and undefinable by proof obligations.

\iffalse 
\input{tab/contract_diagnostics}

\xzw{del the table?} Table~\ref{tab:contract-diagnostics} shows how counterexamples to these obligations reveal differences between contracts.
For example, an input satisfying $P_0\land\neg P$ refutes $L_1$: the reference precondition admits the input, but the generated precondition excludes it.
If the reverse implication $L_3$ also holds, the generated precondition is strictly stronger than the reference one.
\fi 

Figure~\ref{fig:motivating-case} illustrates the corresponding reasoning for postconditions.
The generated contract omits the clause $r\in S$, allowing the illustrated output to violate \iffalse satisfy  $Q\land\neg Q_0$\fi $Q\Rightarrow Q_0$ and refute $L_2$.
The preconditions coincide, and the reference postcondition implies the generated one, so $L_1$, $L_3$, and $L_4$ hold.
The generated contract is therefore complete but unsound, with a strictly weaker postcondition.

\textit{Behavior correctness.}
Following the finite-sample evaluation in Section~\ref{sec:semantics}, we also check contracts against concrete cases.
Here, $I$ denotes the inputs of the target function, including the initial state of mutable parameters.
$O$ denotes its outputs, including the return value and the final state of mutable parameters.
The validated I/O suite contains three kinds of cases.
A correct pair combines an input admitted by the reference precondition with the output that the reference implementation produces for it.
A wrong-output pair combines such an input with an output that the reference contract rejects.
An invalid input either violates the reference precondition or makes the reference implementation panic.
We write $I\models P$ if the input $I$ satisfies $P$, and $O\models Q$ if the output $O$ satisfies $Q$.
Let $\operatorname{Panic}(C,I)$ denote an observed panic when executing the implementation $C$ on input $I$.
The three kinds of cases are checked against the following conditions:
\begin{align}
\text{Correct Pair:}\quad & I\models P \;\land\; O\models Q, \label{eq:pair-acceptance}\\
\text{Wrong-Output Pair:}\quad & I\models P \;\land\; O\models\neg Q, \label{eq:wrong-output-rejection}\\
\text{Invalid Input:}\quad & (I\models\neg P) \;\lor\; \operatorname{Panic}(C,I). \label{eq:invalid-input-rejection}
\end{align}
Both pair conditions require $I\models P$, so a contract cannot reject a wrong output merely by excluding a valid input.
Correct-I/O acceptance, wrong-output rejection, and invalid-input rejection are the fractions of the corresponding cases for which Verus proves these conditions.
An invalid input also counts as rejected if the implementation panics on it.
We record the two separately, because a panic reflects the implementation rather than the contract.
%We record precondition rejection and runtime panic separately, since only the former establishes that the contract excludes the input.

\section{Experimental Setup}
\label{sec:design}
\subsection{Research Questions}
\label{sec:rqs}

\begin{researchquestions}
  \item \textbf{(Quality Proxy):} How well does verifier acceptance reflect specification quality?
  \item \textbf{(Configuration Effects):} How do verifier acceptance and specification quality vary across generation workflows, models, and prompting settings?
  \item \textbf{(Value and Limits of I/O):} What does I/O evaluation add to verifier acceptance, and which contract differences does it miss?
  \item \textbf{(Contract Differences):} Which contract differences do \iffalse the comparison failures\fi the rejected obligations of verifier-accepted artifacts point to, and how can \iffalse they\fi the direction of a failure guide improvements to generated and reference contracts?
\end{researchquestions}

\subsection{Benchmark and Evaluation Datasets}
\label{sec:tasks}

\begin{table}[t]
\caption{Datasets and the RQs that use them.}
\label{tab:datasets}
\centering\footnotesize
\begin{tabular}{@{}llrl@{}}
\toprule
Dataset & Content & Size & RQs \\
\midrule
Tasks & Reference implementation and contract of a target function & 762 & All \\
\quad Labeled & Reference contracts selected by screening and then reviewed & 61 & 4 \\
I/O cases & Correct pairs, wrong-output pairs, and invalid inputs & 11,641 & 1--3 \\
Generated artifacts & 18 configurations: each model zero- and few-shot on all tasks & 13,716 & 2 \\
\quad AlphaVerus & Llama & 1,524 & \\
\quad AutoVerus & GPT-4o & 1,524 & \\
\quad StarVerus & DeepSeek-Chat, DeepSeek-Reasoner, GPT-4o, Llama, Qwen-Coder & 7,620 & \\
\quad VeruSAGE & DeepSeek-Chat, Qwen-Coder & 3,048 & \\
\addlinespace[2pt]
\quad Accepted (eligible) & Accepted by Verus and containing a target function & 7,081 & 1, 3, 4 \\
\quad Proof-failed & Type-valid but failing at proof, with a target function & 1,774 & 1 \\
\quad Flagged & Accepted, with a rejected obligation among $L_1$--$L_4$ & 3,765 & 4 \\
Contract variants & One-side changes of 395 bases equivalent to their references & 943 & 3 \\
\bottomrule
\end{tabular}
\end{table}

\textit{Tasks and references.}
We use the 762 self-contained tasks of StarVerus~\cite{jiang2026starverus} (Table~\ref{tab:datasets}), which originate from VeriCoding~\cite{bursuc2025vericoding} (450), VerusBench~\cite{yang2025autoverus} (144), MBPP-verified~\cite{yang2025autoverus,misu2024dafny} (77), a Dafny-to-Verus collection~\cite{aggarwal2025alphaverus} (53), and HumanEval-Verus~\cite{humanevalverus} (38).
Each task specifies a target function and provides its reference implementation and reference contract $(P_0,Q_0)$, which Verus accepts.

\textit{Validated I/O suites.}
For each task, we build an I/O suite with the three kinds of cases of Section~\ref{sec:implementation}.
An LLM (DeepSeek-V4-Flash) and heuristics for boundary inputs and output mutation propose candidate cases, similar to EvalPlus~\cite{liu2023evalplus}.
A candidate enters the suite only after its label is validated: \iffalse against the reference\fi Verus checks whether its input satisfies the reference precondition and whether the reference postcondition \iffalse contract\fi rejects a wrong output, and running the \iffalse reference\fi implementation supplies correct outputs and confirms panic-based invalid inputs.
The suites thus rest on the same reference as specification correctness.
The suites contain 11,641 cases: 4,716 correct pairs, 4,028 wrong-output pairs, and 2,897 invalid inputs.
Construction targets at least five cases of each kind per task, with lower targets for some tasks, mostly for invalid inputs when a task has no precondition.
Eighty-six tasks fall short for at least one kind.
We evaluate every validated case, and only a kind without cases has no score.

\textit{Generated artifacts.}
We run four generation workflows, AlphaVerus~\cite{aggarwal2025alphaverus}, AutoVerus~\cite{yang2025autoverus}, StarVerus~\cite{jiang2026starverus}, and VeruSAGE~\cite{yang2025verusage}.
StarVerus generates the contract together with the proof code, and we adapt the other three workflows to do the same.
Table~\ref{tab:datasets} lists the 18 configurations of workflow, model, and prompting setting, each run on all 762 tasks.
Llama and Qwen-Coder denote Llama~4 Maverick and Qwen3-Coder-480B-A35B-Instruct.
Of the 13,716 resulting artifacts, Verus accepts 7,134, and the others fail at parsing (1,814), frontend checking (2,990), or proof (1,778).
Because the quality metrics evaluate the target function, we exclude 57 artifacts without an executable one (53 accepted and 4 proof-failed).
The remaining 13,659 artifacts enter the quality metrics of RQ2, and their 7,081 accepted and 1,774 proof-failed members are the two groups that RQ1 compares.

\textit{Controlled contract variants.}
For RQ3, we construct contract variants that differ from the reference in a known direction.
Each variant modifies a base, an accepted artifact that is proved equivalent to its reference and passes a suite with all three kinds of cases.
Of the 7,081 accepted artifacts, 3,937 have such a suite, 2,293 of these pass every case, and 1,430 of those are proved equivalent.
Sampling at most two per task yields 395 bases from 213 tasks.

A variant changes one side of the target contract and keeps the implementation and proof code.
It strengthens the precondition or postcondition by adding a clause (Add) or tightening a comparison, such as $n\geq0$ to $n>0$ (Refine), or weakens it by deleting one of several clauses (Remove) or loosening a comparison, such as $n>0$ to $n\geq0$ (Relax).
A candidate is kept only if Verus still accepts it and confirms the change.
For an original predicate $R$ and a modified predicate $R'$, a stronger $R'$ requires a proof of $R'\Rightarrow R$ and a verified witness of $R\land\neg R'$, and a weaker $R'$ requires the converse.
A second witness shows that a stronger $R'$ remains satisfiable and a weaker $R'$ falsifiable.
We keep the first of up to three passing candidates per base and operation.
Because the base is equivalent to its reference, the variant differs from the reference in the confirmed direction.

Of the 3,552 candidates tried, we keep 943 variants over 204 tasks: 394 stronger and 62 weaker preconditions, and 72 stronger and 415 weaker postconditions.
Weaker preconditions and stronger postconditions are rare because the unchanged implementation must then assume less or guarantee more, so most such candidates fail verification (729 of 845 and 691 of 948).

\subsection{Evaluation Procedure}
\label{sec:protocol}

All evaluations use Verus release 0.2025.09.25.04e8687 with its bundled Rust 1.88.0 toolchain on x86-64 Linux, and each proof attempt has a timeout of 20 seconds.
We retry unresolved attempts with larger time and resource limits and, where needed, with a simplified proof harness that keeps the checked expressions unchanged.
Attempts that still do not complete remain unresolved.
The two intent judgements use DeepSeek-V4-Flash as the judge model.

\section{Experimental Results}
\label{sec:results}
For each RQ, we present the quantitative results and then summarize the answer in a finding.

\subsection{RQ1: Verifier Acceptance as a Quality Proxy}
\label{sec:rq1}

Existing evaluations count every verifier-accepted artifact as a success.
RQ1 examines how well this criterion reflects specification quality.
We first compare accepted and proof-failed artifacts on the quality metrics and then screen the accepted artifacts with three semantic checks.

\textit{Association with quality metrics.}
We compare the 7,081 accepted and 1,774 proof-failed artifacts (Table~\ref{tab:datasets}).
Proof-failed artifacts pass frontend checking but fail at proof, so unlike artifacts that fail earlier, they can be scored on the same metrics as accepted ones.
Table~\ref{tab:acceptance-association} reports the mean of each metric in both groups and Kendall's $\tau_b$~\cite{kendall1938rank,kendall1945ties} between acceptance and the metric, with 95\% confidence intervals from a bootstrap over tasks.
All metrics are oriented so that higher is better, so a positive $\tau_b$ means that accepted artifacts tend to score higher.

% Generated by RQs/RQ1/scripts/refresh_rq1_paper.py.
% Analysis: RQs/RQ1/results/rq1_acceptance_association.csv; sources: outputs_v5/rq1_artifact_labels.csv, outputs_v5/rq1_artifact_outcomes.csv; 2000 task bootstrap replicates; seed 20260826.
\begin{table}[t]
\caption{Quality metrics of accepted and proof-failed artifacts. A dash marks an undefined interval.}
\label{tab:acceptance-association}
\centering\footnotesize
\setlength{\tabcolsep}{3pt}
\begin{tabular}{@{}llrrrrrc@{}}
\toprule
 & & \multicolumn{2}{c}{Accepted} & \multicolumn{2}{c}{Proof-failed} & \multicolumn{2}{c}{$\tau_b$ with acceptance} \\
\cmidrule(lr){3-4}\cmidrule(lr){5-6}\cmidrule(lr){7-8}
Dimension & Metric & $n$ & Mean & $n$ & Mean & Value & 95\% CI \\
\midrule
\multirow{3}{*}{\shortstack[l]{Text\\similarity}} & BLEU & 7,081 & 0.488 & 1,774 & 0.346 & 0.151 & [0.125, 0.178] \\
 & ROUGE-L & 7,081 & 0.655 & 1,774 & 0.519 & 0.174 & [0.149, 0.199] \\
 & KeySpecMatch & 7,081 & 0.715 & 1,774 & 0.633 & 0.112 & [0.082, 0.141] \\
\addlinespace[2pt]
\multirow{2}{*}{\shortstack[l]{Intent\\judgement}} & Spec--code judgement & 7,081 & 0.930 & 1,774 & 0.744 & 0.239 & [0.210, 0.271] \\
 & Spec--spec judgement & 7,081 & 0.778 & 1,774 & 0.588 & 0.196 & [0.165, 0.225] \\
\addlinespace[2pt]
\multirow{2}{*}{\shortstack[l]{Semantic\\triviality}} & False precondition not proved & 7,081 & 0.999 & 1,774 & 1.000 & $-$0.012 & -- \\
 & True postcondition not proved & 7,081 & 0.986 & 1,774 & 0.954 & 0.090 & [0.055, 0.124] \\
\addlinespace[2pt]
\multirow{3}{*}{\shortstack[l]{Specification\\correctness}} & Soundness & 7,074 & 0.694 & 1,771 & 0.379 & 0.261 & [0.225, 0.293] \\
 & Completeness & 7,074 & 0.552 & 1,771 & 0.220 & 0.266 & [0.235, 0.299] \\
 & Equivalence & 7,074 & 0.468 & 1,771 & 0.145 & 0.263 & [0.235, 0.291] \\
\addlinespace[2pt]
\multirow{3}{*}{\shortstack[l]{Behavior\\correctness}} & Correct-I/O acceptance & 6,812 & 0.907 & 1,722 & 0.777 & 0.194 & [0.156, 0.235] \\
 & Wrong-output rejection & 6,592 & 0.958 & 1,660 & 0.810 & 0.239 & [0.201, 0.274] \\
 & Invalid-input rejection & 4,224 & 0.767 & 1,171 & 0.678 & 0.094 & [0.048, 0.142] \\
\bottomrule
\end{tabular}
\end{table}

Acceptance is positively associated with every quality metric except the false-precondition check, which almost no artifact fails.
The association is strongest for specification correctness, with $\tau_b$ between 0.261 and 0.266.
Among the artifacts with an equivalence score, 46.8\% of the accepted and 14.5\% of the proof-failed ones are proved equivalent to their references.
The intent judgements and the two pair-based I/O metrics follow, whereas text similarity, invalid-input rejection, and the true-postcondition check show the weakest associations.

The I/O associations partly reflect unresolved cases rather than incorrect behavior.
For 76 of the 1,774 proof-failed artifacts, Verus reports part of the contract as ill-formed, for example because it cannot infer a trigger for a quantifier.
Most of their case checks therefore remain unresolved and count as not passed.
Excluding these artifacts lowers $\tau_b$ to 0.158 for correct-I/O acceptance and 0.198 for wrong-output rejection, whereas $\tau_b$ for equivalence barely changes (0.257).
The association with specification correctness is thus the most robust.

Acceptance therefore carries information about specification quality but does not certify it, since fewer than half of the accepted artifacts are proved equivalent to their reference contracts, so counting every accepted artifact as a success overstates quality.

\textit{Screening of accepted artifacts.}
A missing equivalence proof, however, does not show that an artifact is wrong.
We therefore screen each accepted artifact with three checks, each of which flags the artifact when it finds evidence of a defect.
\iffalse Reference\fi Contract comparison flags an artifact when Verus rejects one of the implications $L_1$--$L_4$, the I/O suite when the generated contract fails a case, and the triviality check when $T_1$ or $T_2$ is proved.
An artifact passes a check when the check completes without a flag.
Only 2,995 of the 7,081 accepted artifacts (42.3\%) pass all three checks, and 3,790 (53.5\%) are flagged by at least one.
The remaining 296 are not flagged, but at least one of their checks is incomplete, mostly because some I/O cases remain unresolved.
Even counting them as passing would raise the pass rate only to 46.5\%.

\begin{figure}[t]
\centering
\includegraphics{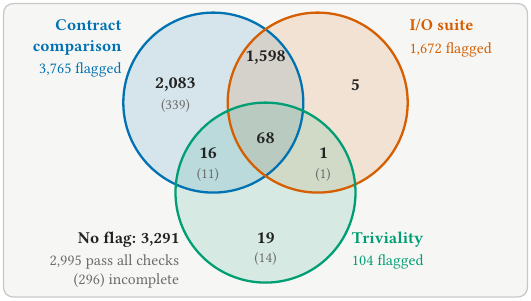}
\caption{Screening flags of the 7,081 accepted artifacts, with counts of incomplete checks in parentheses.}
\Description{The contract-comparison-only, I/O-only, triviality-only, contract-comparison-and-I/O-only, contract-comparison-and-triviality-only, I/O-and-triviality-only, and all-three regions contain 2083, 5, 19, 1598, 16, 1, 68 artifacts. Outside all three circles are 3291 artifacts.}
\label{fig:rq1-check-overlap}
\end{figure}

Figure~\ref{fig:rq1-check-overlap} shows that the three checks are far from equally informative.
Contract comparison also flags 1,666 of the 1,672 I/O-flagged artifacts and 84 of the 104 triviality-flagged ones.
Only 25 flagged artifacts lack a contract-comparison flag, and 22 of them are proved equivalent to their reference contracts.
For 17 of these 22, the triviality obligation proved for the generated contract therefore also holds for the reference contract.
The I/O suite thus rarely detects a problem that contract comparison misses.

The flags, however, differ in strength.
A failed case usually provides a concrete counterexample, and a proved triviality obligation shows that the contract is vacuous.
A rejected implication, by contrast, shows only that a proof attempt failed.
Where the I/O suite and contract comparison flag the same artifact, the failed case usually supplies the counterexample that the rejected implication lacks.
Overall, 44.7\% of the flagged artifacts have a counterexample or a proved triviality obligation.
The 2,083 artifacts (55.0\%) flagged by contract comparison alone rest only on rejected implications.

\finding{1}{Verifier acceptance is positively associated with specification quality, most robustly with specification correctness, but it overstates quality as a success criterion.
Only 42.3\% of the accepted artifacts pass all three semantic checks, and 53.5\% are flagged by at least one.
Of the flagged artifacts, 44.7\% have a counterexample or a proved triviality obligation, whereas 55.0\% rest only on rejected implications.}

\subsection{RQ2: Effects of Generation Configurations}
\label{sec:rq2}

RQ1 shows that acceptance is an imperfect proxy for the quality of individual artifacts.
RQ2 asks whether a configuration with higher acceptance also produces better specifications.
We vary one factor at a time.
Workflows are compared on a shared model and models within StarVerus, both under few-shot prompting, and prompting settings are compared within each StarVerus model.

\begin{figure}[t]
\centering
\includegraphics[width=\linewidth]{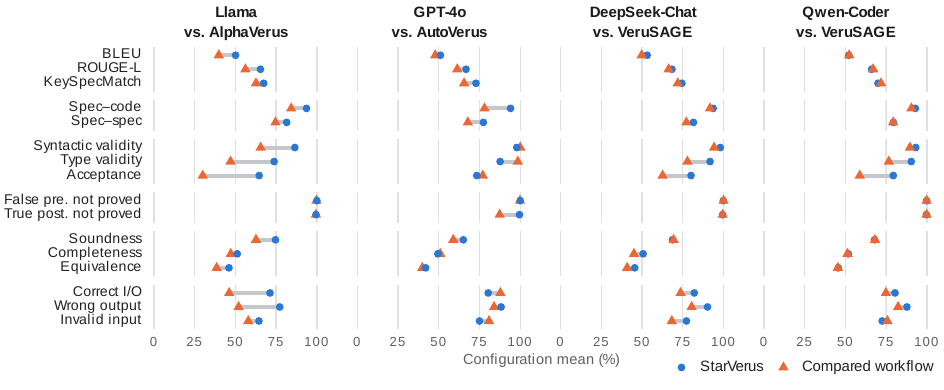}
\caption{Workflows compared on a shared model under few-shot prompting.}
\Description{Four panels compare StarVerus with AlphaVerus using Llama, AutoVerus using GPT-4o, and VeruSAGE using DeepSeek-Chat and Qwen-Coder. Sixteen metrics are grouped by the six dimensions. Blue circles show StarVerus and orange triangles the compared workflow. Verifier validity uses all 762 tasks per configuration; quality means exclude artifacts without an executable target and use the available scores. Workflow rankings differ across metrics.}
\label{fig:workflow-comparison}
\end{figure}

\begin{figure}[t]
\centering
\includegraphics[width=\linewidth]{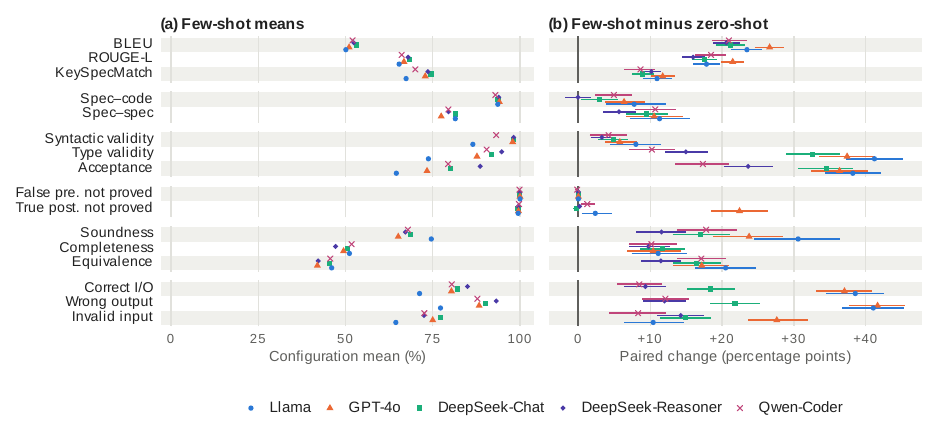}
\caption{StarVerus models: (a) few-shot means and (b) paired zero- to few-shot changes with 95\% intervals.}
\Description{Two aligned panels show sixteen metrics for five StarVerus models, distinguished by color and shape. Panel (a) shows configuration means over the available scores, with quality metrics restricted to artifacts with an executable target. Panel (b) shows paired changes using the same scored tasks for each model and metric. Acceptance, correct-I/O acceptance, and wrong-output rejection increase for all five models; their intervals lie above zero.}
\label{fig:model-prompt-comparison}
\end{figure}

Figures~\ref{fig:workflow-comparison} and~\ref{fig:model-prompt-comparison}(a) show configuration means.
Unlike RQ1, which compares only artifacts that reach the proof stage, RQ2 averages each quality metric over all artifacts of a configuration that have a score, including those that fail parsing or frontend checking.
Most I/O cases of such artifacts remain unresolved and count as not passed, so the I/O means also reflect how often a configuration produces type-valid artifacts.
Since the scored artifacts differ between configurations, a difference in means may partly reflect which tasks are scored.
For the differences we discuss, we therefore also compare the two configurations on the tasks that both have scored, with 95\% bootstrap intervals over tasks.
Figure~\ref{fig:model-prompt-comparison}(b) shows such paired changes from zero-shot to few-shot prompting.

% Source: RQs/RQ2/results/rq2_configuration_profiles.csv; plot_rq2_figures.R.

\textit{Workflow differences.}
Across workflows, the acceptance gap does not predict the quality gap.
In two of the four comparisons, the workflow with higher acceptance also leads on most metrics: StarVerus over AlphaVerus with Llama and over VeruSAGE with DeepSeek-Chat.
In the other two, it does not.
With GPT-4o, AutoVerus accepts more tasks than StarVerus (77.0\% versus 73.4\%) but scores lower on the spec--code judgement (78.1\% versus 94.1\%) and the true-postcondition check (87.4\% versus 99.6\%).
With Qwen-Coder, StarVerus accepts 20.6 percentage points more tasks than VeruSAGE, yet the two differ by at most 0.5 points in soundness, completeness, and equivalence.

% Sources: RQs/RQ2/results/rq2_configuration_profiles.csv and rq2_prompt_deltas.csv; plot_rq2_figures.R.

\textit{Model differences.}
Across models, acceptance rankings carry over to the pair-based I/O metrics but not to specification correctness.
Within StarVerus, acceptance ranges from 64.6\% for Llama to 88.6\% for DeepSeek-Reasoner.
Type validity follows the same ordering, and so, largely, do the two pair-based I/O metrics, which depend on it.
Specification correctness does not.
On the tasks scored for both models, DeepSeek-Reasoner accepts 24.0 points more tasks than Llama but shows no clear difference in soundness, completeness, or equivalence, with all intervals including zero.
It is even 3.2--4.3 points lower than DeepSeek-Chat and Qwen-Coder in completeness and equivalence.

\textit{Prompting differences.}
Prompting, in contrast, moves acceptance and quality in the same direction.
Few-shot prompting raises acceptance for all five StarVerus models by 17.3--38.2 points and improves most quality metrics, with larger gains in soundness than in completeness.
Much of the I/O gain, however, reflects artifacts that become type-valid, so that their cases can be checked, rather than contracts that behave better.
Few-shot prompting raises type validity most for Llama and GPT-4o (by 41.2 and 37.4 points), which are also the two models with the largest I/O gains.
On tasks where both settings yield type-valid artifacts, the few-shot gain drops to at most 6.2 points for correct-I/O acceptance and 9.0 for wrong-output rejection.

\finding{2}{Within the comparisons of this study, four workflow pairs and the model and prompting
comparisons in StarVerus, acceptance is an unreliable basis for comparing configurations: \iffalse across workflows and models,\fi higher acceptance does not consistently come with higher specification correctness.
Few-shot prompting raises acceptance and most quality metrics, although much of its I/O gain reflects artifacts that become type-valid.}

\subsection{RQ3: Coverage and Limits of I/O Evaluation}
\label{sec:rq3}

RQ1 and RQ2 show that verifier acceptance must be complemented by semantic checks.
Our validated I/O suites provide the finite-sample evidence of Section~\ref{sec:semantics}, and nearly all I/O-flagged artifacts are also flagged by \iffalse cases labeled by the same reference as\fi contract comparison (Figure~\ref{fig:rq1-check-overlap}).
RQ3 asks what this evidence adds to verifier acceptance and where it falls short of contract comparison.
We measure two gaps on the accepted artifacts: the gap from acceptance to I/O passing shows the usefulness of I/O evaluation, and the gap from I/O passing to proved soundness, completeness, and equivalence shows its limitation.
We then evaluate the controlled variants of Section~\ref{sec:tasks}, whose differences from the reference contract are known, to explain which differences the finite cases miss.
Because the suites and contract comparison share the oracle $(P_0,Q_0)$, their agreement is not independent corroboration, so the variant experiment measures the suites themselves.

\textit{From acceptance to I/O passing and equivalence.}
We start from the 7,081 accepted artifacts in Figure~\ref{fig:rq1-check-overlap}.
The I/O suite flags 1,672 of them (23.6\%), which acceptance alone would count as successes, and 660 others have unresolved or no cases.
The remaining 4,749 (67.1\%) pass every available case.
Among these, however, only 3,737 (78.7\%) are proved sound, 3,373 (71.0\%) complete, and 3,000 (63.2\%) equivalent to their references.
Table~\ref{tab:rq3-conditional} shows that this gap persists in every configuration, with equivalence rates between 51.0\% and 70.5\% among I/O-passing artifacts.

% Generated by RQs/RQ3/scripts/refresh_rq3_paper.py from results/natural/conditional_rates.csv.
\begin{table}[t]
\caption{Accepted artifacts (Acc.), \% passing I/O, and \% of those proved equivalent (Equiv.) by configuration.}
\label{tab:rq3-conditional}
\centering\footnotesize
\begin{tabular}{@{}llrrrrrr@{}}
\toprule
 & & \multicolumn{3}{c}{Zero-shot} & \multicolumn{3}{c}{Few-shot} \\
\cmidrule(lr){3-5}\cmidrule(lr){6-8}
Workflow & Model & Acc. & I/O & Equiv. & Acc. & I/O & Equiv. \\
\midrule
\multirow{1}{*}{AlphaVerus} & Llama & 42 & 33.3 & 57.1 & 224 & 74.1 & 70.5 \\
\addlinespace[2pt]
\multirow{1}{*}{AutoVerus} & GPT-4o & 197 & 55.3 & 67.0 & 587 & 69.0 & 65.4 \\
\addlinespace[2pt]
\multirow{5}{*}{StarVerus} & DeepSeek-Chat & 347 & 66.9 & 60.3 & 610 & 70.0 & 64.9 \\
 & DeepSeek-Reasoner & 493 & 60.6 & 59.9 & 672 & 67.4 & 60.9 \\
 & GPT-4o & 261 & 57.1 & 66.4 & 558 & 69.0 & 62.9 \\
 & Llama & 200 & 56.5 & 51.3 & 487 & 70.4 & 67.3 \\
 & Qwen-Coder & 472 & 62.5 & 53.9 & 590 & 71.7 & 64.1 \\
\addlinespace[2pt]
\multirow{2}{*}{VeruSAGE} & DeepSeek-Chat & 166 & 69.3 & 69.6 & 477 & 69.6 & 67.8 \\
 & Qwen-Coder & 250 & 62.0 & 51.0 & 448 & 74.6 & 66.2 \\
\midrule
All &  & 2,428 & 61.0 & 59.1 & 4,653 & 70.2 & 65.0 \\
\bottomrule
\end{tabular}
\end{table}

% Generated by RQs/RQ3/scripts/refresh_rq3_paper.py from results/variants/.
\begin{table}[t]
\caption{Detection of controlled variants by kind of case. Five variants yield only expression diagnostics.}
\label{tab:io-detection}
\centering\footnotesize
\setlength{\tabcolsep}{4pt}
\begin{tabular}{@{}lllrrrrrr@{}}
\toprule
Change & Violates & Operation & $n$ & Correct & Wrong & Invalid & Detected & Missed \\
\midrule
\multirow{2}{*}{\shortstack[l]{Stronger\\precondition}} & \multirow{2}{*}{$L_1$} & Add & 163 & 161 & 128 & 0 & 161 & 2 \\
 &  & Refine & 231 & 230 & 227 & 0 & 230 & 1 \\
\addlinespace[2pt]
\multirow{2}{*}{\shortstack[l]{Weaker\\precondition}} & \multirow{2}{*}{$L_3$} & Remove & 35 & 0 & 0 & 27 & 27 & 8 \\
 &  & Relax & 27 & 0 & 0 & 19 & 19 & 8 \\
\addlinespace[2pt]
\multirow{2}{*}{\shortstack[l]{Stronger\\postcondition}} & \multirow{2}{*}{$L_4$} & Add & 69 & 0 & 0 & 0 & 0 & 69 \\
 &  & Refine & 3 & 0 & 0 & 0 & 0 & 3 \\
\addlinespace[2pt]
\multirow{2}{*}{\shortstack[l]{Weaker\\postcondition}} & \multirow{2}{*}{$L_2$} & Remove & 224 & 0 & 163 & 0 & 163 & 58 \\
 &  & Relax & 191 & 0 & 171 & 0 & 171 & 18 \\
\midrule
Total &  &  & 943 & 391 & 689 & 46 & 771 & 167 \\
\bottomrule
\end{tabular}
\end{table}

\textit{Detection by direction.}
We run the 943 variants on the suites of their bases (Table~\ref{tab:io-detection}).
Each kind of case can witness only one direction of a variant's change: \iffalse whose counterexample in Table~\ref{tab:contract-diagnostics} it can satisfy.\fi a case with input satisfying $P_0$ (a correct or wrong-output pair) witnesses a stronger precondition ($P_0\land\neg P$), an invalid input violating $P_0$ witnesses a weaker precondition ($P\land\neg P_0$), a correct output satisfying $Q_0$ witnesses a stronger postcondition ($Q_0\land\neg Q$), and a wrong output outside $Q_0$ witnesses a weaker postcondition ($Q\land\neg Q_0$).
\iffalse A stronger postcondition requires an output in $Q_0\land\neg Q$: wrong outputs violate $Q_0$ and thus also the stronger postcondition, so only a correct pair whose reference output the new postcondition rejects can expose it.\fi 

In practice, the suites detect 771 of the 943 variants (81.8\%), miss 167 that pass every case, and give only expression diagnostics for the other five.
Correct pairs detect 391 of the 394 stronger preconditions, invalid inputs 46 of the 62 weaker preconditions, and wrong-output pairs 334 of the 415 weaker postconditions.
Wrong-output pairs also detect 355 stronger preconditions, because their check requires the input to be admitted first.
In the current suites, no case detects any of the 72 stronger postconditions.
%This zero is structural, not empirical. A retained variant keeps the implementation and the precondition of its base and is still \textit{accepted} by Verus, so the implementation satisfies the stronger postcondition wherever the reference admits the input.

\textit{Why changes escape the suite.}
A change escapes when no case in the suite is judged differently by the original and the changed contract.
For 95 of the 167 misses, which change a precondition (19) or weaken a postcondition (76), the suite simply lacks such a case.
In one task, for example, deleting the precondition clause $N>0$ newly admits inputs with $N=0$, but every invalid input in the suite also violates the remaining clause such as $|a|=N$ and is still rejected.
The 72 stronger postconditions escape under this construction. A retained variant keeps the implementation and the precondition of its base and is still \textit{accepted} by Verus, so the implementation satisfies the stronger postcondition wherever the reference admits the input. This explanation is conditional on the base implementation agreeing with the reference implementation on the reference domain that generated the positive pairs; we state it as an assumption about this retained suite, not as a general implementation-equivalence theorem.
\iffalse 
A stronger postcondition rejects only outputs that the original one accepts.
The wrong outputs in a suite are rejected by both contracts, and its correct outputs come from the reference implementation, which satisfies the stronger postcondition in all 72 cases.
For example, one variant adds the guarantee that a vector keeps its length.
The original contract also allows other lengths, but the reference implementation always keeps the length, so no case distinguishes the two contracts.
Such a guarantee is not necessarily wrong, since it describes what the implementation does where the reference contract leaves the behavior open.
\fi 

\finding{3}{I/O evaluation adds screening beyond acceptance but is weak evidence of equivalence.
It rejects 23.6\% of the accepted artifacts, yet only 63.2\% of those that pass it are proved equivalent to their references, and every configuration shows this gap.
On 943 controlled contract variants, the suites detect 81.8\%, but detection depends on the direction of the change.
\iffalse They miss all 72 stronger postconditions, because the only permitted outputs in the suites come from the reference implementation and satisfy these postconditions.\fi 
Under the construction and the stated base/reference behavioral-agreement assumption, the 72 stronger postconditions are expected to escape; the suites detect 88.5\% of the other 871 variants.
}

\subsection{RQ4: Contract Differences and Improvement Directions}
\label{sec:rq4}

RQ1 and RQ3 leave contract comparison as the most discriminating check: it flags 3,765 of the 7,081 accepted artifacts, including nearly all that the I/O suite flags.
Its flags, however, only identify rejected proof obligations, not confirmed differences.
The percentages below therefore describe proof-obligation patterns and screening priorities, not a confirmed distribution of semantic defects.
RQ4 asks which contract differences these rejections point to and how they can guide improvements to generated and reference contracts.
We call a rejected obligation among $L_1$--$L_4$ a failure, with the direction given in Equations~\eqref{eq:l1}--\eqref{eq:l4}.
\iffalse A failure alone does not confirm a difference from the reference contract.
When an artifact fails only one obligation, however, Verus proves the other three, so any difference can lie only in the failed direction.\fi 

\begin{figure}[t]
\centering
\includegraphics[width=\linewidth]{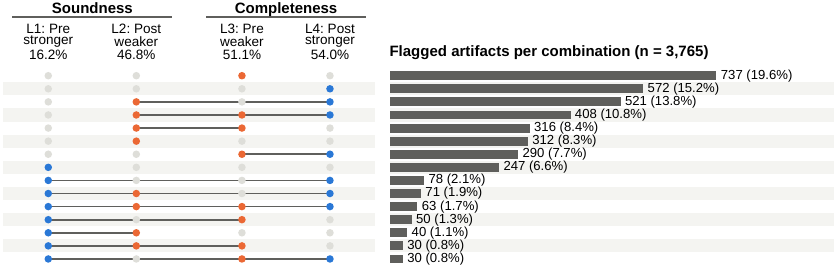}
\caption{Combinations of failed obligations in the 3,765 flagged artifacts.}
\Description{An UpSet chart laid out by rows, with combinations sorted by frequency. The columns group L1, stronger precondition, and L2, weaker postcondition, under Soundness, and L3, weaker precondition, and L4, stronger postcondition, under Completeness. Filled dots mark rejected obligations; all unmarked obligations are proved. Column shares are 16.2, 46.8, 51.1, and 54.0 percent. Bars give the number and share of artifacts with each combination. Direction labels indicate potential differences requiring confirmation.}
\label{fig:rq4-relations}
\end{figure}

\textit{Failure directions.}
Figure~\ref{fig:rq4-relations} shows the failures of the 3,765 flagged artifacts, which cover 634 tasks.
Most of them have a rejected completeness obligation: 84.1\% have $L_3$ or $L_4$ rejected, compared with 57.5\% that have a rejected soundness obligation through $L_1$ or $L_2$.
For preconditions, weaker-precondition failures ($L_3$, 51.1\%) far outnumber stronger ones ($L_1$, 16.2\%).
Postcondition failures are common in both directions ($L_4$ 54.0\%, $L_2$ 46.8\%).
The most frequently rejected obligation is thus $L_4$, the stronger-postcondition direction in which the I/O suites of RQ3 detect none of the variants.

%\xzw{Because $L_1$ and $L_3$ are necessary for the corresponding inclusions whereas $L_2$ and $L_4$ are not (Section 3.2), rejections of $L_1$ and $L_3$ are more likely to reflect real differences, while rejections of $L_2$ and $L_4$ may additionally stem from comparing postconditions outside the preconditions’ domain.}
%(the min_array repair below shows one such spurious $L_2$ rejection)

\textit{Contract differences.}
The column shares count each obligation separately, whereas each row of Figure~\ref{fig:rq4-relations} shows which obligations fail together and which are proved.
The proved obligations determine how the generated contract relates to its reference.
If all rejected obligations of a row lie under Completeness, $L_1$ and $L_2$ are proved, so the generated contract is sound and thus (equivalent to or) stronger \iffalse not weaker\fi than the reference (Section~\ref{sec:semantics}).
These rows cover 1,599 flagged artifacts (42.5\%), about half of the 3,166 with a rejected completeness obligation.
Conversely, the rows whose rejected obligations all lie under Soundness cover 599 artifacts (15.9\%), whose generated contracts are complete and thus (equivalent to or) weaker \iffalse not stronger\fi than the reference.
In another 571 artifacts (15.2\%), the preconditions (521, rejecting only $L_2$ and $L_4$) or the postconditions (50, rejecting only $L_1$ and $L_3$) are proved equivalent.
The difference thus lies in the other condition, which may be weaker in some respects and stronger in others, so both cases below can apply to it.
The remaining 996 artifacts (26.5\%) establish none of these relations.

\textit{Completeness proved, soundness unproved.}
For the 599 complete artifacts, a rejected $L_1$ suggests a precondition that excludes inputs the reference admits, and a rejected $L_2$ a postcondition that permits outputs the reference rejects.
Both differences can be exposed by I/O cases (Table~\ref{tab:io-detection}).
A correct pair whose input the generated precondition excludes shows an input to admit, and a wrong output that the generated postcondition accepts shows an output to exclude.
Such counterexamples indicate how to revise the generated contract, by relaxing its precondition or tightening its postcondition.

\textit{Soundness proved, completeness unproved.}
For the 1,599 sound artifacts, a rejected $L_3$ suggests a precondition that admits inputs the reference excludes, and a rejected $L_4$ a postcondition that guarantees more than the reference.
A confirmed difference of either kind would make the contract stronger, and the accepted artifact proves that its implementation handles the extra inputs or provides the extra guarantees.
Executing the implementation thus cannot yield a counterexample to a stronger postcondition, and although an invalid input may expose the extra inputs of a weaker precondition, the implementation is proved correct on them too.
Such differences therefore prioritize reviewing the reference, while a stronger generated contract may also overconstrain the task: the reference may abstract the behavior on purpose, restrict inputs the task allows, or omit a guarantee that the task requires.
We do not claim that the generated contract is correct, only that it is Verus-verified and equivalent to or stronger than the reference. A stronger contract from a single model is thus only a candidate, and recurrence across models signals which references warrant review, as implemented below.

\textit{Reviewing reference contracts.}
We screen the reference contracts with five signals, each of which selects a reference for review against its implementation.
S0 applies the triviality obligations $T_1$ and $T_2$ to the reference itself, since any generated contract that is not itself vacuous fails completeness against a vacuous reference.
S1 and S2 look for the same problems through accepted artifacts, selecting a task with a postcondition proved true ($T_2$) that still implies the reference postcondition ($L_2$) or with a precondition proved false ($T_1$).
S3 and S4 look for comparison patterns shared across models.
S3 selects a task with at least six accepted artifacts, at least 80\% of which, from at least three models, have $L_2$ proved and $L_4$ rejected, the pattern of the stronger guarantees above.
S4 applies the same rule to preconditions ($L_3$ proved, $L_1$ rejected).

% Generated data: RQs/RQ5/results/v5/screen_signals.csv and screen_tasks.csv; plot: RQs/RQ4/scripts/refresh_rq4_rq5_paper.py.
\begin{figure}[t]
\centering
\includegraphics[width=\linewidth]{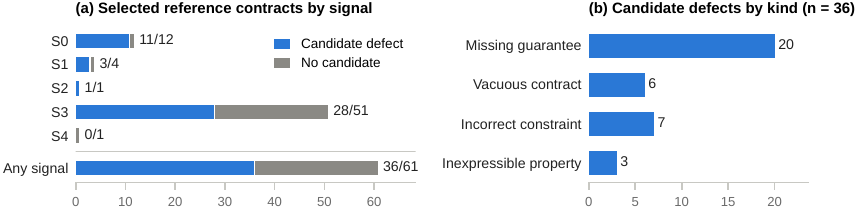}
\caption{Screening of reference contracts: (a) candidates/selected per signal and (b) candidate defects by kind.}
\Description{Two horizontal bar charts. Panel (a) lists the signals S0 to S4 and any signal. Each bar is split into candidate defects in blue and references without a candidate label in gray. S0 selects 12 references with 11 candidates, S1 4 with 3, S2 1 with 1, S3 51 with 28 candidates, S4 1 with none, and any signal 61 with 36 candidates and 25 without a candidate label. Panel (b) shows the 36 candidate defects by kind: 20 missing guarantees, 6 vacuous contracts, 7 incorrect constraints, and 3 inexpressible properties.}
\label{fig:rq5-screen}
\end{figure}

\begin{figure}[t]
\centering
\includegraphics[width=\linewidth]{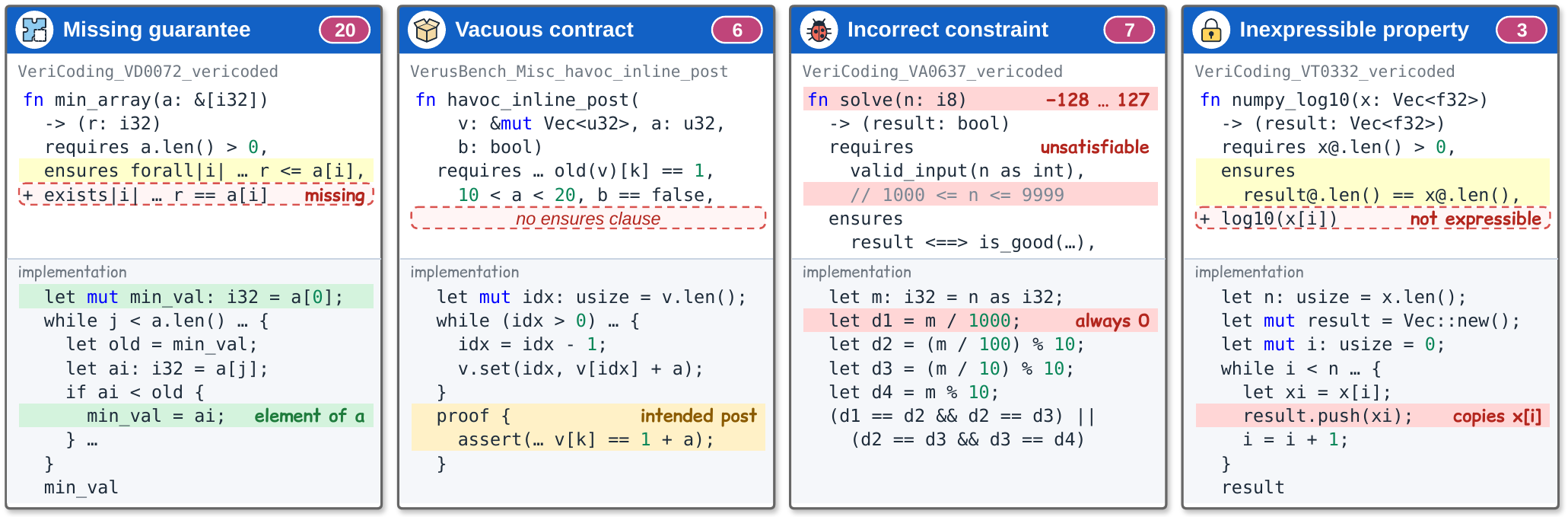}
\caption{One candidate reference defect per kind. Badges give counts, and dashed boxes mark omissions.}
\Description{Four cards, one per kind of candidate reference defect, each showing a reference contract above an implementation excerpt. Missing guarantee, 20 references: the min_array reference of VeriCoding VD0072 ensures only that the result is at most every element. A dashed box marks the missing clause that the result equals some element, and the implementation indeed returns an element of the array. Vacuous contract, 6 references: the havoc_inline_post reference has no ensures clause, and the intended postcondition, that every element equals 1 plus a, appears only as an assertion in a proof block. Incorrect constraint, 7 references: the reference of VeriCoding VA0637 takes an i8 parameter, whose values range from -128 to 127, but requires it to lie between 1000 and 9999, so the precondition is unsatisfiable, and the implementation's leading digit m / 1000 is always 0. Inexpressible property, 3 references: the numpy_log10 reference of VeriCoding VT0332 ensures only that the result has the input length, because the base-10 logarithm of f32 values is not expressible, and the implementation copies each input element.}
\label{fig:rq5-defects}
\end{figure}

\textit{Reference review results.}
The signals select a non-random, purposive sample of 61 of the 762 references (8.0\%, Figure~\ref{fig:rq5-screen}a); the resulting 36/61 is a yield among screened references, not an estimate of benchmark-wide defect prevalence.
Two independent judges assess whether each selected reference captures the observable behavior of its implementation, that is, its return value and the final state of the parameters it writes, and identify the candidate defects.
They agree on 57 of the 61, and a domain expert then reviews every label to finalize it, including adjudication of all four disagreements.
The five signals, their selection thresholds, the independent judgments, and expert adjudication define the screening sample; references outside it are not represented by this rate.
%Because this review compares references with implementations rather than with confirmed task requirements, we call the defects it finds candidates.
The 36 candidate defects fall into four kinds (Figure~\ref{fig:rq5-screen}b).
A missing guarantee leaves part of the behavior unconstrained (20), a vacuous contract admits no input or constrains no output (6), an incorrect constraint misstates a condition, such as a value range (7), and an inexpressible property cannot be stated in a Verus contract (3).
Figure~\ref{fig:rq5-defects} shows one candidate of each kind.

% Built by RQs/RQ5/scripts/build_reference_defects_figure.py; excerpts from the hash-checked reference sources.

\textit{Repairing a reference contract.}
The first example in Figure~\ref{fig:rq5-defects} illustrates one repair selected from the screened references; its 10/11 result is a case study, not an estimate of repair success across the benchmark.
All 11 accepted artifacts of \texttt{min\_array}, from five models, fail only $L_4$ because they require the minimum to be an array element, a guarantee the reference omits and whose loss a comment attributes to translation from Dafny.
Their shared guarantee thus suggests the repair.
We add it to the reference with the loop invariants that prove it, and Verus accepts the repaired reference with the unchanged implementation.
Against it, 10 of the 11 artifacts are proved equivalent, and the remaining one fails $L_2$ only because the postcondition comparison does not assume $|a|>0$.

\finding{4}{The direction of a comparison failure prioritizes inspection of the generated contract or the reference; it does not by itself establish which artifact is defective.
Of the 3,765 flagged artifacts, 15.9\% fail only soundness obligations, a pattern for inspecting weaker generated contracts and using I/O counterexamples to guide relaxed preconditions or tightened postconditions.
Another 42.5\% fail only completeness obligations, a pattern that prioritizes auditing the reference, while a stronger generated contract may still overconstrain the task.
Reviewing the 61 screened references finds 36 candidate defects, and a guarantee shared by several models can repair a reference that omits it.}

\section{Actionable Lessons}
\label{sec:lessons}
Our findings yield four lessons for evaluation, generation, and benchmark maintenance.

\textit{Lesson 1: Report contract quality alongside acceptance.}
Acceptance alone overstates the quality of generated contracts (Finding~1) and misleads comparisons, because a workflow or model with higher acceptance does not necessarily produce more correct contracts (Finding~2).
Evaluations should therefore report triviality, specification correctness, and behavior correctness together with acceptance, and compare configurations on the same tasks.
Few-shot prompting raises both acceptance and quality for the StarVerus models, but much of its I/O gain comes from artifacts that become type-valid, so I/O gains should be reported together with type validity.

\textit{Lesson 2: Design I/O suites for the differences they must expose.}
An I/O suite reveals a contract change only if one of its cases is judged differently by the original and the changed contract (Finding~3).
Our suites show two gaps.
First, a weakened constraint goes unnoticed when every invalid input also violates another constraint that remains.
Invalid inputs should therefore each violate a different input constraint, and wrong outputs a different output guarantee.
Second, no case in the current suites detects the 72 stronger postconditions, which are expected to escape under the construction and the stated base/reference behavioral-agreement assumption (Section~\ref{sec:rq3}).
Such a difference need not be a defect of the generated contract, and Lesson~4 handles it through the reference rather than the suite.
Evaluations should also report the three kinds of cases separately, since a wrong-output case fails not only when the contract accepts the wrong output but also when a stronger precondition rejects the input.
For invalid inputs, they should distinguish precondition rejection from runtime panic, which can conceal an omitted input constraint.

\textit{Lesson 3: Revise weaker generated contracts with I/O counterexamples.}
Weaker generated contracts may exclude inputs that the reference admits or permit outputs that it rejects, and the 15.9\% pattern prioritizes inspecting this possibility (Finding~4).
Verifier feedback does not reveal such differences, since the artifacts are accepted.
I/O cases do, as the RQ3 variants show (Table~\ref{tab:io-detection}).
Each failing case shows what to change, an input to admit or an output to exclude.
Generation workflows should therefore feed failing cases back to the generator and re-verify the revised contract.
Such cases can come from task requirements or existing tests when no reference is available.

\textit{Lesson 4: Review the reference when stronger contracts recur across models.}
The 42.5\% pattern prioritizes reviewing stronger generated contracts and the references against which they are compared (Finding~4).
Their extra inputs or guarantees are met by the verified implementation, so \iffalse such a difference questions the reference rather than the generated contract\fi a confirmed difference of this kind makes the reference a candidate for review rather than proving the generated contract wrong, although a stronger contract may also overconstrain the task.
When the same difference recurs across models, the reference may omit a guarantee that the task requires, and screening can select it for review.
For \texttt{min\_array}, adding the guarantee that the generated contracts share repairs the reference and makes 10 of its 11 flagged artifacts equivalent (Section~\ref{sec:rq4}).
Benchmark maintainers should therefore review such references before treating the generated contracts as wrong.
A shared guarantee should be adopted only if the task requires it and the unchanged implementation verifies against the extended reference, and the I/O labels should be revalidated afterwards.

\textit{Summary.}
Together, the lessons suggest a staged evaluation.
Acceptance serves as an entry check, and the triviality checks remove vacuous contracts.
With a reference, the direction of a comparison failure prioritizes the next inspection: weaker generated contracts can be tested and revised with I/O counterexamples, while stronger ones that recur across models can trigger a review of the reference.
Without a reference contract, I/O cases of all three kinds give partial evidence, and a stronger postcondition that the implementation satisfies passes every case.

\section{Threats to Validity}
\label{sec:threats}
\textit{Construct validity.}
Contract comparison and the I/O suites both depend on the reference, which can itself be defective.
RQ4 screens the reference contracts for such defects and reviews a non-random sample of 61 selected ones; the 36/61 result is a screened-sample yield, references outside the selection are not reviewed, and the rate of undetected defects is unknown.
A rejected obligation may also stem from the limits of proof automation rather than a real difference, so we treat it as a failed proof attempt rather than a confirmed difference.

\textit{Internal validity.}
Errors in extracting contracts or building proof harnesses could distort results.
We encode every obligation natively in Verus, which avoids translation into another logic.
The RQ3 variants come from a limited, retained set of edits; their detection rates are descriptive for these mutations and do not estimate detection of naturally occurring defects.
The reference review of RQ4 compares contracts with implementations, not task requirements, so we report candidate defects rather than a confirmed defect rate.
Each configuration generates one artifact per task, so configuration comparisons are based on one generation draw per task; the bootstrap over tasks captures task variation but not run-to-run variation.

\textit{External validity.}
Our 762 self-contained tasks may not represent repository-scale verification~\cite{liu2026vostd}, although they come from five benchmarks.
Not every model runs in every workflow, so we compare workflows only on shared models.
All results use one Verus release (Section~\ref{sec:protocol}), and another release may decide some obligations differently.

\section{Related Work}
\label{sec:related-work}
\textbf{Proof Generation.}
LLMs support deductive verification by generating proof annotations and repairing failed proofs.
AutoVerus~\cite{yang2025autoverus} synthesizes and repairs Verus proof annotations using verifier feedback, while VeruSAGE~\cite{yang2025verusage} studies agentic proof completion for Rust systems.
RagVerus~\cite{zhong2025ragverus} retrieves repository context for proof generation, and VerusSeek~\cite{zhang2026verusseek} applies retrieval to Rust proof synthesis.
ExVerus~\cite{yang2026exverus} uses counterexample reasoning to guide Verus proof repair.
With a supplied contract, verifier acceptance measures whether the generated proof code completes verification.

\noindent\textbf{Specification Generation.}
Specification generators infer the contracts that state the properties to be verified.
SpecGen~\cite{ma2025specgen} combines conversational specification generation with mutation of candidate specifications to obtain verifiable contracts.
AutoSpec~\cite{wen2024autospec} uses static analysis and program verification to decompose programs and iteratively validate generated specifications.
Fun2Spec~\cite{ugare2026fun2spec} synthesizes quantified contracts for C++ programs using automated validation and error feedback.
For unsafe Rust, KaPilot~\cite{wang2026kapilot} extracts safety requirements from documentation, refines Kani specifications through verification, and uses implication checks to select among candidates.
For Verus, AlphaVerus~\cite{aggarwal2025alphaverus} generates code and contracts through translation and refines candidates using verifier feedback.
StarVerus~\cite{jiang2026starverus} uses multiple agents to generate specifications and proofs.
These approaches make the contract itself an output that verification alone cannot validate.

\noindent\textbf{Specification Evaluation.}
Empirical studies examine how well generated specifications capture program behavior.
Xie et al.~\cite{xie2025effective} compare LLMs and traditional techniques for generating specifications from comments and documentation, studying few-shot prompting and generation failures.
Endres et al.~\cite{endres2024postconditions} evaluate generated postconditions by their correctness and ability to discriminate incorrect code.
EvoSpex~\cite{molina2021evospex} learns postconditions from valid and invalid program states, and SpecSyn~\cite{ma2026specsyn} uses program mutants to assess and refine specification strength.
Recent evaluation frameworks use tests, proofs, and consistency checks to assess generated specifications.
Verus-SpecGym~\cite{agarwal2026verusspecgym} executes Verus specifications against official contest tests and adversarial cases.
Coins~\cite{yang2026coins} instantiates specifications on trusted test cases to produce Rocq proof obligations and highlights how verification difficulty can obscure specification quality.
Clover~\cite{sun2024clover} checks consistency among code, docstrings, and formal annotations.
LiveFMBench~\cite{xu2026livefmbench} shows that prover gaming and ignored code context can inflate measured specification-generation accuracy.

Our controlled variants and reference screening draw on work on test adequacy and benchmark quality.
Mutation analysis~\cite{jia2011mutation} seeds small faults to measure whether a test suite detects them, and EvalPlus~\cite{liu2023evalplus} shows that insufficient tests overstate the correctness of generated code.
Label errors in test sets can likewise change model rankings on machine learning benchmarks~\cite{northcutt2021labelerrors}.
Our work complements these studies in a common Verus setting.
It relates verifier acceptance, I/O cases, and logical comparison with the reference on the same artifacts and measures which contract differences each can reveal.
It also uses the direction of a comparison failure to prioritize inspection of the generated contract or the reference, while treating the reference itself as a possible source of error and using differences shared across models to find and repair candidate defects.

\section{Conclusion}
\label{sec:conclusion}
We presented VerusEval, a framework that evaluates LLM-generated Verus contracts along six dimensions, and applied it to 13,716 artifacts on 762 tasks.
Our study revealed that verifier acceptance overstates contract quality: only 42.3\% of the accepted artifacts \iffalse are proved equivalent to their references and pass both the I/O suite and the triviality checks,\fi  complete all three semantic checks without a flag, and across workflows and models, higher acceptance does not consistently come with more correct contracts.
I/O evaluation rejected 23.6\% of the accepted artifacts but was weak evidence of equivalence.
\iffalse and it missed all 72 controlled variants that strengthen the postcondition.\fi 
Most accepted artifacts flagged by contract comparison have a rejected completeness obligation,
\iffalse that fail contract comparison fail completeness,\fi  and the direction of a failure helps prioritize inspection of the generated contract or the reference but does not by itself identify which artifact is defective.
Screening and review identified 36 candidate defects among the benchmark reference contracts, and repairing one of them made 10 of its 11 failing artifacts equivalent.
We derived four actionable lessons for evaluation, generation, and benchmark maintenance.
These findings highlight the need to evaluate generated contracts not only by whether they verify but also by what they guarantee, and to treat benchmark references as subject to review.

% FSE 2027: mandatory, placed after Conclusion, does not count toward the
% page limit. Keep author-anonymous (anonymous artifact link only).
% \section{Data Availability}
% \label{sec:data-availability}
% \input{sec/10_data_availability}

\bibliographystyle{ACM-Reference-Format}
\bibliography{sample-base}

\end{document}